\documentclass[letterpaper, preprint, paper,11pt]{AAS}	
\usepackage[utf8]{inputenc}
\usepackage{amsmath, amsfonts, overcite,float}
\usepackage{caption}
\usepackage{subcaption}

\PaperNumber{22-752}

\newcommand{\R}{\mathbb{R}}

\begin{document}

\title{Orbital Acceleration Using Product of Exponentials}

\author{Taylor Yow\thanks{Masters Student, Space Technologies Laboratory, Embry-Riddle Aeronautical University, 1 Aerospace Blvd, Daytona Beach, FL., 32114},  
Christopher W. Hays\thanks{PhD Candidate, Space Technologies Laboratory, Embry-Riddle Aeronautical University, 1 Aerospace Blvd, Daytona Beach, FL., 32114},
Aryslan Malik\thanks{Visiting Assistant Professor, Aerospace Engineering Department, Space Technologies Laboratory,Embry-Riddle Aeronautical University, 1 Aerospace Blvd, Daytona Beach, FL., 32114},
and Troy Henderson\thanks{Associate Professor, Aerospace Engineering Department, Space Technologies Laboratory, Embry-Riddle Aeronautical University, 1 Aerospace Blvd, Daytona Beach, FL., 32114}
}

\maketitle{} 		

\begin{abstract}
The Product of Exponentials (PoE) formulation is most commonly used in the field of robotics, but has recently been adapted for use in describing orbital motion. The PoE formula for orbital mechanics is an alternate method for defining and drawing an orbit based on its orbital elements set. Currently the PoE formula for orbital mechanics has only been derived through the first derivative (velocity). This work explores the second derivative of the adapted PoE formula for orbital mechanics, which gives a more complete description of the orbital motion of a satellite in a two-body system. This comprehensive approach employs a unified approach to account for all six time-varying orbital elements, therefore broadening the scope of the research and applications.
\end{abstract}

\section{Introduction}

Brockett's study was the first to adopt the Product of Exponentials (PoE) formulation to express forward kinematics of a multi-body system \cite{POE1}. The PoE formulation's core concept is to consider a joint as a screw that moves the rest of the external links (bodies), which allows for simultaneous rotation and translation resulting in a more concise mathematical formulation \cite{POE2,lynch2017modern,tsiotras1995new,holm2009geometricmechanics,poe3, poe4}. The PoE has proven to be useful for robotic control and manipulation \cite{interest1,poe_robotics1,interest2,poe_robotics2,interest3,poe_robotics3,interest4,poe_robotics4,interest5,poe_robotics5,interest6,poe_robotics6,interest7,interest8,interest9}.

It was conceptualized by Malik\cite{malik2022using} to apply the PoE formulation to ``drawing" orbits where each term in the orbital element set $[\Omega,\ i,\ \omega,\ \theta,\ r(\theta)]$ acts as a separate joint.
This allows for many of the tools available within PoE for robotic arms to also be applied to the natural motion of orbital mechanics. 
However, there is a notable lack of representation of acceleration within the PoE framework, which has mainly focused on robotic arm kinematics. 
In the orbital regime, acceleration plays a critical role in the description and characterization of motion through Newton's Second Law.
For this reason, the present work focuses on extending the PoE framework to characterize acceleration specifically tailored for the orbital motion arena.

\section{Drawing Orbits with Product of Exponentials}

The PoE-orbital mechanics framework relates a set of orbital elements $([\Omega, i, \omega, \theta, r(\theta)])$ to the state (position, velocity) of the satellite $(p_{eo}\in{}\R^3,{V}_{eo}\in{}\R^3)$ using exponential mappings \cite{malik2022using}. Note that $\Omega$ is the right ascension of the ascending node, $i$ is the inclination of the orbit, $\omega$ is the argument of periapsis, $\theta$ is the true anomaly, and $r(\theta)$ is the location of the spacecraft defined as: 
\begin{equation}\label{rtheta}
r(\theta)=\frac{a(1-e^2)}{1+e\cos\theta}
\end{equation}
where $a$ is the semi-major axis and $e$ is the eccentricity. The position of a body frame expressed in the inertial frame is calculated here leveraging a single PoE formula:
\begin{equation}\label{eq:bodyframe}
{T_{eo}}=e^{[\mathcal{S}_1]\Omega}e^{[\mathcal{S}_2]i}e^{[\mathcal{S}_3]\omega}e^{[\mathcal{S}_4]\theta}e^{[\mathcal{S}_5]r(\theta)}M_{eo}\in{}SE(3)
\end{equation}
where, 
\begin{equation}
M_{eo}=\begin{bmatrix}
           1&0&0&0\\
           0&1&0&0\\
           0&0&1&0\\
           0&0&0&1\\ \end{bmatrix}=I_{4\times{}4}
\end{equation}

\begin{equation} \label{S_matrices}
    \mathcal{S}_{i} = 
    \begin{bmatrix}
    w \\
    u
    \end{bmatrix};\ 
    \mathcal{S}_1 = 
    \begin{bmatrix}
    0 \\0\\1\\0\\0\\0
    \end{bmatrix}
    \mathcal{S}_2 =
    \begin{bmatrix}
    1\\0\\0\\0\\0\\0
    \end{bmatrix}
    \mathcal{S}_3 = 
    \begin{bmatrix}
    0\\0\\1\\0\\0\\0
    \end{bmatrix}
    \mathcal{S}_4 = 
    \begin{bmatrix}
    0\\0\\1\\0\\0\\0
    \end{bmatrix}
    \mathcal{S}_5 = 
    \begin{bmatrix}
    0\\0\\0\\1\\0\\0
    \end{bmatrix}
\end{equation}
Configuration matrices $T_{eo}$ are part of $SE(3)$, which is the Special Euclidean space that preserves the Euclidean distance between any two points and is described by $4\times{}4$ matrices that store both attitude and position information as shown in Equation \ref{explanation1}. The subscript $(eo)$ denotes that $M_{eo}$ is the ``home" configuration of an ``object" relative to the equatorial frame of the major body. The $M_{eo}$ is the $4\times{}4$ identity matrix because initially it is assumed that the object's position and orientation coincides with the the equatorial frame's position and orientation, which is then modified by exponentials which allow the transformation of our object's frame. 
\begin{figure}[h!] 
    \centering
    \includegraphics[scale=0.6]{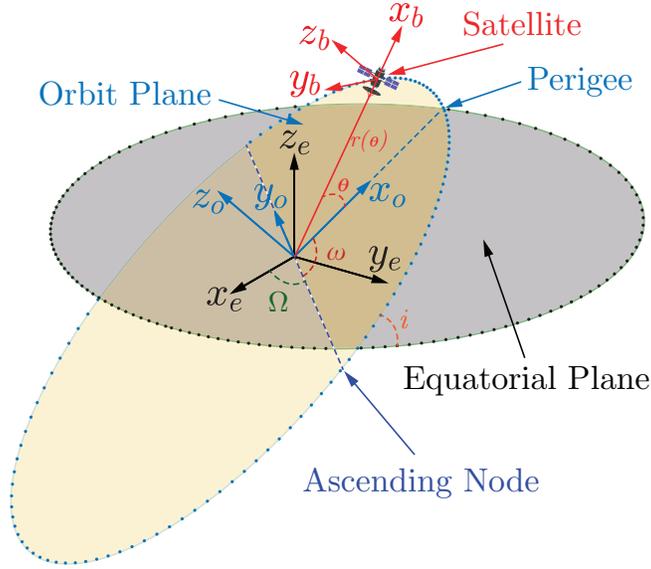}
    \caption{Representation of the body frame of the satellite.}
    \label{bodyframe}
\end{figure}

In Figure \ref{bodyframe}, $(x_e,\,y_e,\,z_e)$ is the geocentric equatorial frame, $(x_o,\,y_o,\,z_o)$ is the orbital perifocal frame, and $(x_b,\,y_b,\,z_z)$ is the satellite body frame with $x_b$ pointing radially outwards, $y_b$ is in the direction of the velocity of the satellite, and $z_b$ is normal to the orbital plane.

The inertial velocity of an object in orbit expressed in an inertial frame  is given by the following equation:
\begin{equation}\label{velocitybody}
 {V}_s=R_{eo}\begin{bmatrix}
           \dot{r}(\theta)\\
           r(\theta)\dot{\theta}\\
           0\\
           \end{bmatrix}=R_{eo}{V}_b
\end{equation}
where $R_{eo}\in{}SO(3)$ is extracted from the configuration of the body,
\begin{equation}\label{explanation1}
{T_{eo}}=\begin{bmatrix}
           R_{eo} & p_{eo}\\
           0 & 1 \\
            \end{bmatrix}\in{}SE(3)
\end{equation}
More generally, the inertial velocity of the body is calculated directly from $\mathcal{V}_s\in\R^6$ spatial twist as follows:
\begin{align}\label{11}
    \mathcal{V}_s&=\mathcal{S}_1\dot{\Omega}+\mathrm{Ad}_{e^{[\mathcal{S}_1]\Omega}}(\mathcal{S}_2)\dot{i}+\mathrm{Ad}_{e^{[\mathcal{S}_1]\Omega}e^{[\mathcal{S}_2]i}}(\mathcal{S}_3)\dot{\omega}\\&+\mathrm{Ad}_{e^{[\mathcal{S}_1]\Omega}e^{[\mathcal{S}_2]i}e^{[\mathcal{S}_3]\omega}}(\mathcal{S}_4)\dot{\theta}+\mathrm{Ad}_{e^{[\mathcal{S}_1]\Omega}e^{[\mathcal{S}_2]i}e^{[\mathcal{S}_3]\omega}e^{[\mathcal{S}_4]\theta}}(\mathcal{S}_5)\dot{r}(\theta)\nonumber
\end{align}
where,
\begin{equation}\label{vs1}
    \mathcal{V}_s=\begin{bmatrix}
           \omega_s\\
           v_s\\
           \end{bmatrix}\in\R^6
\end{equation}
and where,
\begin{equation}\label{eq: radial_velocity}
    \dot{r}(\theta) = \frac{dr(\theta)}{d\theta} \frac{d\theta}{dt} = \frac{dr(\theta)}{d\theta}\dot{\theta} = \frac{ae(1-e^2)\sin{}\theta}{(1+e\cos\theta)^2}\dot{\theta}=e\sin{}\theta\sqrt{\frac{\mu}{a(1-e^2)}}
\end{equation}
as,
\begin{equation}\label{eq:thetadoteq}
    \dot{\theta}=\frac{\sqrt{\mu{}a(1-e^2)}}{r^2}=(1+e\cos\theta)^2\sqrt{\frac{\mu}{[a(1-e^2)]^3}}
\end{equation}

The adjoint operator is also defined as,
\begin{equation}\label{eq: adjoint_definition}
    \mathrm{Ad}_{T} = 
    \begin{bmatrix}
    R & 0_{3\times 3}\\
    p^{\times}R & R
    \end{bmatrix}
\end{equation}
with $(\cdot)^{\times}:| \mathbb{R}^{3}\rightarrow \mathfrak{so}(3)$ being the skew-symmetric operator maps the vector space to the Lie algebra, defined in Equation \ref{eq: skew_symmetric}.
\begin{equation}\label{eq: skew_symmetric}
    p^{\times} = 
    \begin{bmatrix}
    0 & -p_3 & p_2\\
    p_3 & 0 & -p_1\\
    -p_2 & p_1 & 0
    \end{bmatrix}
\end{equation}
and the $[\hat{\cdot}]:|\mathbb{R}^{6}\rightarrow \mathfrak{se}(3)$ also maps the vector space to its corresponding Lie algebra and is defined in Equation \ref{eq: R6_Lie_algebra}.
\begin{equation}\label{eq: R6_Lie_algebra}
    [\hat{\mathcal{V}_{s}}] = 
    \begin{bmatrix}
    \omega_s^{\times} & v_s \\
    0_{1\times3} & 0
    \end{bmatrix}
\end{equation}
Therefore, the velocity vector of the satellite expressed in the inertial frame becomes:
\begin{equation}\label{inertial_velocity}
    {V}_s=v_s+\omega_s\times{p_{eo}}\in\R^3
\end{equation}
A current limitation of the PoE formulation is it is only defined to the velocity level. 
Which means in its current form, PoE cannot accommodate external forces and control torques impressed upon the body.
For a more generalized concept, this work directly extends the PoE framework to the acceleration level, which will allow for explicit consideration of external and control forces.

\section{Development of Acceleration Representation}

\subsection{Direct Derivative}
By taking the derivative with respect to time of the spatial twist expressed in Equation \ref{11}, the spatial acceleration in Equation \ref{eq: poe_acceleration} can be obtained.
\begin{multline}\label{eq: poe_acceleration}
   \mathcal{A}_s = \mathrm{Ad}_{\dot{T_1}}(S_2)\dot{i} + \mathrm{Ad}_{T_1}(S_2)\ddot{i} + \mathrm{Ad}_{\dot{T_1} T_2}(S_3)\dot{\omega} +\mathrm{Ad}_{T_1 \dot{T_2}}(S_3)\dot{\omega} + \mathrm{Ad}_{T_1 T_2}(S_3)\ddot{\omega} + \mathrm{Ad}_{\dot{T_1} T_2 T_3}(S_4)\dot{\theta} \\
   + \mathrm{Ad}_{T_1 \dot{T_2} T_3}(S_4)\dot{\theta}+\mathrm{Ad}_{T_1 T_2 \dot{T_3}}(S_4)\dot{\theta} + \mathrm{Ad}_{T_1 T_2 T_3}(S_4)\ddot{\theta} + \mathrm{Ad}_{\dot{T_1} T_2 T_3 T_4}(S_5)\dot{r}(\theta)\\
   + \mathrm{Ad}_{T_1 \dot{T_2} T_3 T_4}(S_5)\dot{r}(\theta) + \mathrm{Ad}_{T_1 T_2 \dot{T_3} T_4}(S_5)\dot{r}(\theta) + \mathrm{Ad}_{T_1 T_2 T_3 \dot{T_4}}(S_5)\dot{r}(\theta) + \mathrm{Ad}_{T_1 T_2 T_3 T_4}(S_5)\ddot{r}(\theta)
\end{multline}
In an effort to simplify notation, each of the exponential terms were rewritten as,
\begin{equation}\label{eq: compact_definitions}
    \begin{split}
         T_1 &= e^{[S_1]\Omega}\\
         T_2 &= e^{[S_2]i} \\
         T_3 &= e^{[S_3]\omega}  \\
         T_4 &= e^{[S_4]\theta}\\
         T_5 &= e^{[S_5]r(\theta)}\\
    \end{split}
\end{equation}
From Modern Robotics\cite{lynch2017modern} the exponential matrices can be expanded into the form of Equation \ref{eq: MR_exponential}. The form of Equations \ref{eq: MR_exponential} and \ref{eq: G_exponential} are the same for all S matrices in Equation \ref{S_matrices}.
\begin{equation}\label{eq: MR_exponential}
    e^{[S_1]\Omega} = \begin{bmatrix}
        e^{[\omega]\Omega} & G(\Omega)v\\
        0 & 1\\
    \end{bmatrix}
\end{equation}
where
\begin{equation}\label{eq: G_exponential}
    G(\Omega) = I\Omega+(1-cos{\Omega})[\omega]+(\Omega-sin{\Omega})[\omega]^2
\end{equation}

It is also important to note the existence of the second derivative of true anomaly and the second derivative of the scalar range term.
\begin{equation}\label{eq:thetaddoteq}
    \ddot{\theta}=(-2e\sin{\theta})(1+e\cos\theta)\sqrt{\frac{\mu}{[a(1-e^2)]^3}}\dot{\theta}
\end{equation}
\begin{equation}\label{eq: radial_acceleration}
    \ddot{r}(\theta) = e\cos\theta\sqrt{\frac{\mu}{a(1-e^2)}}\dot{\theta}
\end{equation}

 Through the Adjoint action composition law \cite{holm2011geometric}, shown in Equation \ref{eq: adjoint_composition}, the Adjoint operator is distributive and allows Equation \ref{eq: poe_acceleration} to be expanded into Equation \ref{eq: poe_acceleration_expanded} for easier substitution later on.

 \begin{equation}\label{eq: adjoint_composition}
    \mathrm{Ad}_{{T_1}{T_2}{T_3}{T_4}} = \mathrm{Ad}_{T_1}\mathrm{Ad}_{T_2}\mathrm{Ad}_{T_3}\mathrm{Ad}_{T_4}
\end{equation}
\begin{multline}\label{eq: poe_acceleration_expanded}
          \mathcal{A}_s = \mathrm{Ad}_{\dot{T_1}}(S_2)\dot{i} + \mathrm{Ad}_{T_1}(S_2)\ddot{i} + \mathrm{Ad}_{\dot{T_1}}\mathrm{Ad}_{T_2}(S_3)\dot{\omega} +\mathrm{Ad}_{T_1}\mathrm{Ad}_{\dot{T_2}}(S_3)\dot{\omega} + \mathrm{Ad}_{T_1}\mathrm{Ad}_{ T_2}(S_3)\ddot{\omega} \\+ \mathrm{Ad}_{\dot{T_1}}\mathrm{Ad}_{T_2}\mathrm{Ad}_{T_3}(S_4)\dot{\theta} + \mathrm{Ad}_{T_1}\mathrm{Ad}_{ \dot{T_2}}\mathrm{Ad}_{T_3}(S_4)\dot{\theta}+\mathrm{Ad}_{T_1}\mathrm{Ad}_{T_2}\mathrm{Ad}_{\dot{T_3}}(S_4)\dot{\theta} + \mathrm{Ad}_{T_1}\mathrm{Ad}_{T_2}\mathrm{Ad}_{T_3}(S_4)\ddot{\theta}\\+ \mathrm{Ad}_{\dot{T_1}}\mathrm{Ad}{T_2}\mathrm{Ad}_{T_3}\mathrm{Ad}_{T_4}(S_5)\dot{r}(\theta)+ \mathrm{Ad}_{T_1}\mathrm{Ad}_{\dot{T_2}}\mathrm{Ad}_{T_3}\mathrm{Ad}_{ T_4}(S_5)\dot{r}(\theta) + \mathrm{Ad}_{T_1}\mathrm{Ad}{T_2}\mathrm{Ad}_{\dot{T_3}}\mathrm{Ad}_{T_4}(S_5)\dot{r}(\theta) \\+ \mathrm{Ad}_{T_1} \mathrm{Ad}_{T_2}\mathrm{Ad}_{T_3}\mathrm{Ad}_{\dot{T_4}}(S_5)\dot{r}(\theta) + \mathrm{Ad}_{T_1}\mathrm{Ad}_{T_2}\mathrm{Ad}_{T_3}\mathrm{Ad}_{ T_4}(S_5)\ddot{r}(\theta)
\end{multline}
Up until this point, the focus of the paper is been placed on taking the direct derivative of the velocity term from Equation \ref{11} to form the acceleration in Equation \ref{eq: poe_acceleration_expanded} with little attention paid to the key insights within the adjoint operator. 
The following section will describe in more detail the adjoint operator, its derivatives and their constructions.


\subsection{Constructing Adjoint Matrices}
Recalling the construction of the $T\in SE(3)$ elements specified in Equation \ref{explanation1} and the definitions for $T_{i}\ \forall\  i=1,\dots,4$ from Equation \ref{eq: compact_definitions},  $R_1$ and $R_2$ are obtained through Rodrigues' Rotation Formula
\begin{equation}
    R_1 = I + \sin\Omega w^{\times}_{S_1} + (1-\cos\Omega)w^{\times}_{S_1}{}^2
\end{equation}
\begin{equation}
    R_2 = I + \sin{}iw^{\times}_{S_2} + (1-\cos{}i)w^{\times}_{S_2}{}^2
\end{equation}

Where $[\hat{w}]_{S_n}$ is the first three elements of the screw axis ($S$). The last three elements of the screw axis are denoted as $V_{S_n}$, which will be used later on.
\begin{equation}\label{eq: rot_1}
    R_1 = 
    \begin{bmatrix}
    \cos\Omega & -\sin\Omega & 0 \\ \sin\Omega & \cos\Omega & 0\\ 0&0&1
    \end{bmatrix}
\end{equation}
\begin{equation}\label{eq: rot_2}
    R_2 = 
    \begin{bmatrix}
   1&0&0\\ 0 & \cos i & -\sin i \\0& \sin i & \cos i
    \end{bmatrix}
\end{equation}
Since the screw axes $S_3$ and $S_4$ are the same as $S_1$, $R_3$ and $R_4$ are of the same form as $R_1$.
\begin{equation}\label{eq: rot_3}
    R_3 = 
    \begin{bmatrix}
    \cos\omega & -\sin\omega & 0 \\ \sin\omega & \cos\omega & 0\\ 0&0&1
    \end{bmatrix}
\end{equation}
\begin{equation}\label{eq: rot_4}
    R_4 = 
    \begin{bmatrix}
    \cos\theta & -\sin\theta & 0 \\ \sin\theta & \cos\theta & 0\\ 0&0&1
    \end{bmatrix}
\end{equation}

Substituting Equations \ref{eq: rot_1}, \ref{eq: rot_2}, \ref{eq: rot_3}, \ref{eq: rot_4} into Equation \ref{eq: adjoint_definition} yields the rotational matrix components of the $\mathrm{Ad}_{T_{i}}$ terms for Equation \ref{eq: poe_acceleration_expanded}.
At this point it is necessary to point out that
 $p$ in all of these cases is a zero vector, and therefore $[\hat{p}]$ is a zero matrix.
 As a result, each of the adjoint matrices are constructed as shown in Equation \ref{eq: constructed_adjoints}
\begin{equation}\label{eq: constructed_adjoints}
    \begin{split}
    \mathrm{Ad}_{T_1} &= 
    \begin{bmatrix}
    \cos\Omega & -\sin\Omega&0&0&0&0\\\sin\Omega &\cos\Omega&0&0&0&0\\0&0&1&0&0&0\\0&0&0&\cos\Omega & -\sin\Omega&0\\0&0&0&\sin\Omega &\cos\Omega&0\\0&0&0&0&0&1
    \end{bmatrix}\\
      \mathrm{Ad}_{T_2} &= 
    \begin{bmatrix}
    1&0&0&0&0&0\\0&\cos{}i&-\sin{}i&0&0&0\\0&\sin{}i&\cos{}i&0&0&0\\0&0&0&1&0&0\\0&0&0&0&\cos{}i&-\sin{}i\\0&0&0&0&-\sin{}i&\cos{}i
    \end{bmatrix}\\
     \mathrm{Ad}_{T_3} &= 
    \begin{bmatrix}
    \cos{}\omega & -\sin{}\omega&0&0&0&0\\\sin{}\omega &\cos{}\omega&0&0&0&0\\0&0&1&0&0&0\\0&0&0&\cos{}\omega & -\sin{}\omega&0\\0&0&0&\sin{}\omega &\cos{}\omega&0\\0&0&0&0&0&1
    \end{bmatrix}\\
     \mathrm{Ad}_{T_4} &= 
    \begin{bmatrix}
    \cos{}\theta & -\sin{}\theta&0&0&0&0\\\sin{}\theta &\cos{}\theta&0&0&0&0\\0&0&1&0&0&0\\0&0&0&\cos{}\theta & -\sin{}\theta&0\\0&0&0&\sin{}\theta &\cos{}\theta&0\\0&0&0&0&0&1
    \end{bmatrix}
       \end{split}
\end{equation}

\subsection{$\mathbf{SE(3)}$-based Kinematics}

The on-manifold kinematics are given by Equation \ref{eq: se3_kinematics}.
Because each of the screw axes, $\mathcal{S}_{n}$ are unit vectors, the element is included to scale the vectors by the appropriate magnitude.
The full kinematic description of the body configuration is given by
\begin{equation}\label{eq: se3_kinematics}
    \dot{T}_{eo} = 
    \begin{bmatrix}
    \hat{\mathcal{V}_{s}}
    \end{bmatrix}T_{eo}\\
\end{equation}
However in the context of Equation \ref{eq: poe_acceleration_expanded}, only the individual rotational kinematics shown in Equation \ref{eq: compact_definitions} are of interest.
The rotational kinematics of each of these elements are calculated by taking the derivative of Equation \ref{eq: compact_definitions},
\begin{equation}\label{eq: compact_kinematic_definitions}
    \begin{split}
         \dot{T}_1 &= \dot{\Omega}[S_1]e^{[S_1]\Omega}\\
         \dot{T}_2 &= \dot{i}[S_2]e^{[S_2]i} \\
         \dot{T}_3 &= \dot{\omega}[S_3]e^{[S_3]\omega}  \\
         \dot{T}_4 &= \dot{\theta}[S_4]e^{[S_4]\theta}\\
    \end{split}
\end{equation}

The derivative of the Adjoint of an element is just the Adjoint of the derivative of the element, as defined in Equation \ref{eq: adjoint_derivative}.  
\begin{equation}\label{eq: adjoint_derivative}
    \mathrm{Ad}_{\dot{T_n}} = 
    \begin{bmatrix}
    \dot{R_n} & 0_{3x3}\\ 0_{3x3}& \dot{R_n}
    \end{bmatrix}
\end{equation}

\subsection{Acceleration from Spatial Twist}
The acceleration can be also computed from the spatial twist by taking the inertial derivative of the position vector $(p_{eo})$ twice, which yields:
\begin{equation}\label{eq:inertial_acceleration_spatial_twist}
    {A}_s=\frac{\Ddot{r}(\theta)}{r(\theta)}p_{eo}+\dot{\omega}_s\times{}p_{eo}+\frac{2\dot{r}(\theta)}{r(\theta)}\omega_s\times{}p_{eo}+\omega_{s} \times (\omega_{s} \times p_{eo})
\end{equation}
where the angular acceleration vector of the body frame $\dot{\omega}_s$ can be obtained from the spatial twist derivative (spatial acceleration):
\begin{equation}\label{eq:acc_alt}
    \mathcal{A}_s=\dot{\mathcal{V}}_s=\begin{bmatrix}
           \dot{\omega}_s\\
           \dot{v}_s\\
           \end{bmatrix}\in\R^6,
\end{equation}
or from the spatial twist directly:
\begin{equation}\label{eq:wsdot}
\dot{\omega}_s=-2\frac{V_s\cdot{}p_{eo}}{r^2(\theta)}\omega_s=-2\frac{[v_s+\omega_s\times{p_{eo}}]\cdot{}p_{eo}}{r^2(\theta)}\omega_s
\end{equation}
Using this alternative formulation Equations \ref{rtheta}, \ref{11}, \ref{eq: radial_velocity}, \ref{inertial_velocity}, \ref{eq: radial_acceleration}, and \ref{eq:wsdot} are substituted into Equation \ref{eq:inertial_acceleration_spatial_twist} to describe the acceleration of a body in orbit.
\section{Results}

\begin{figure}[] 
    \centering
    \begin{subfigure}[b]{0.49\textwidth}
        \centering
        \includegraphics[width=1\textwidth]{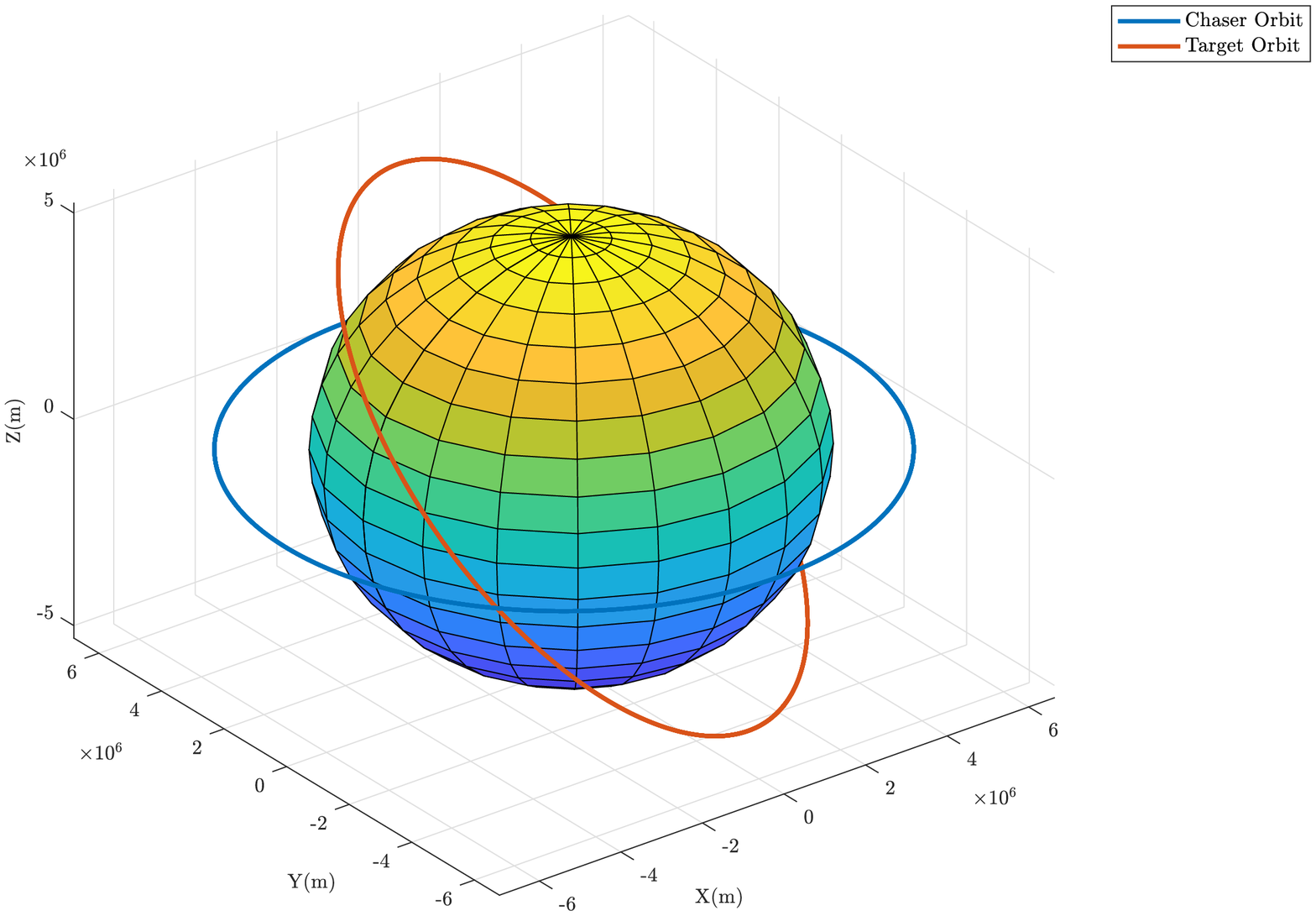}
        \caption{Isometric view.}
        \label{testorbit}
    \end{subfigure}
    \hfill
    \begin{subfigure}[b]{0.49\textwidth}
        \centering
        \includegraphics[width=1\textwidth]{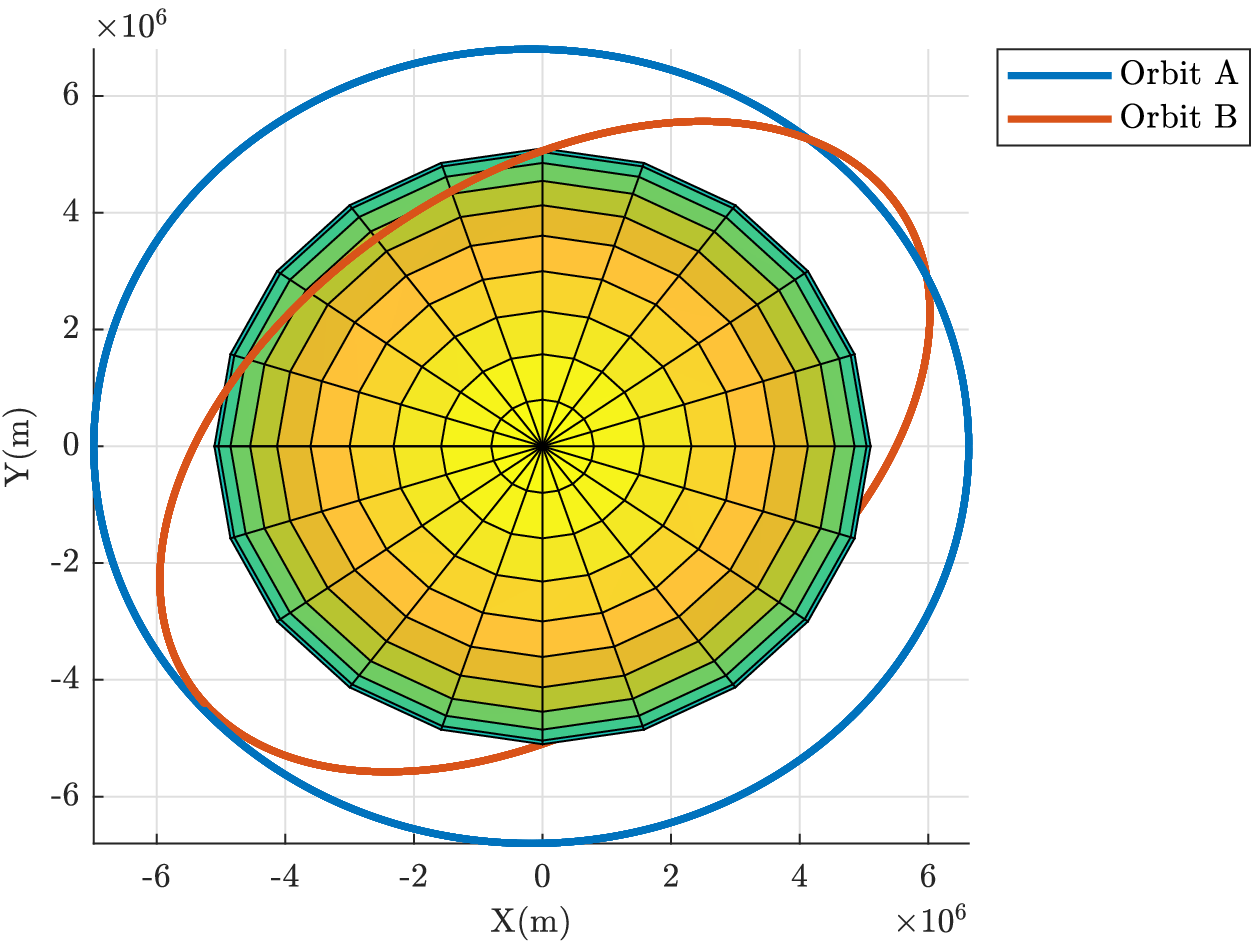}
        \caption{Top view.}
        \label{testorbit}
    \end{subfigure}
  \caption{Representation of Test orbits A and B about Earth.}
  \label{fig:testorbits}
\end{figure}

To validate this method of using PoE to calculate an orbital acceleration it was compared to traditional orbital acceleration equations using two orbits - a very basic elliptical orbit and a circular orbit. The details of both orbits are presented in Table \ref{tab:orbits}, and their visualization is shown in Figure \ref{fig:testorbits}.
\begin{table}[htbp]
	\fontsize{10}{10}\selectfont
    \caption{Description of Orbits}
   \label{tab:orbits}
        \centering 
   \begin{tabular}{c | r | r } 
      \hline 
       Orbital Element & Orbit A & Orbit B\\
      \hline 
 $a$      & $13E6$ km   & $13E6$ km    \\ 
 $e$      & $0.3$        & $0.0073$         \\
 $i$      & $0^{\circ}$ & $50^{\circ}$  \\
 $\omega$ & $0^{\circ}$  & $120^{\circ}$   \\
 $\Omega$ & $0^{\circ}$  & $40^{\circ}$   \\
      \hline
   \end{tabular}
\end{table}

The following equation demonstrates the relationship between PoE acceleration and classical acceleration. 
\begin{equation}\label{eq: classical acceleration}
    A_s = \ddot{p}_{eo}= -{\mu}\frac{p_{eo}}{|p_{eo}|^3} 
\end{equation}
In Figure 3 it can be seen that the accelerations calculated using the PoE method are the same as the accelerations calculated using classical orbital mechanics for both test orbits. The relationship in Equation \ref{eq: classical acceleration} was proven correct with an error being the machine precision. 
\begin{figure}[H] 
    \centering
    \begin{subfigure}[b]{0.49\textwidth}
        \centering
        \includegraphics[width=1\textwidth,trim={0 0 0.8cm 0},clip]{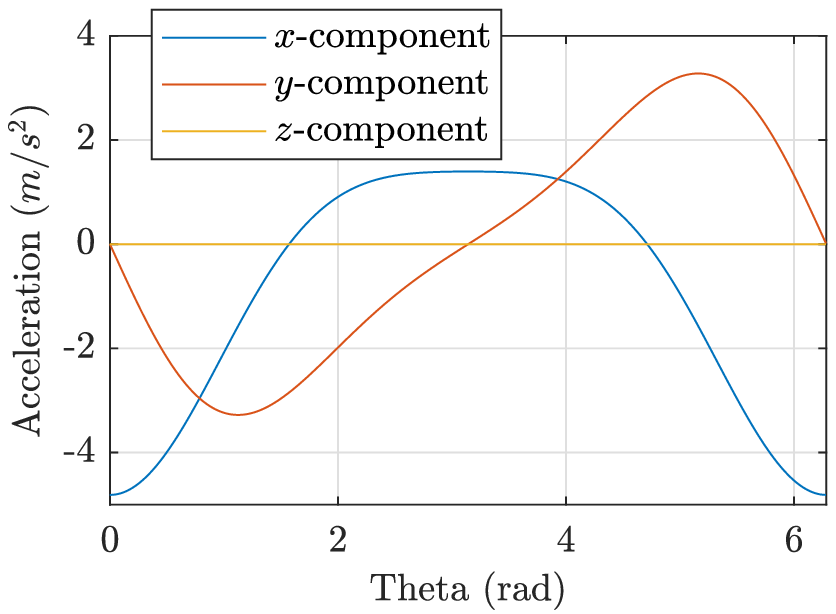}
        \caption{Orbit A acceleration plot.}
        \label{classicalA}
    \end{subfigure}
    \hfill
    \begin{subfigure}[b]{0.49\textwidth}
        \centering
        \includegraphics[width=1\textwidth,trim={0 0 0.8cm 0},clip]{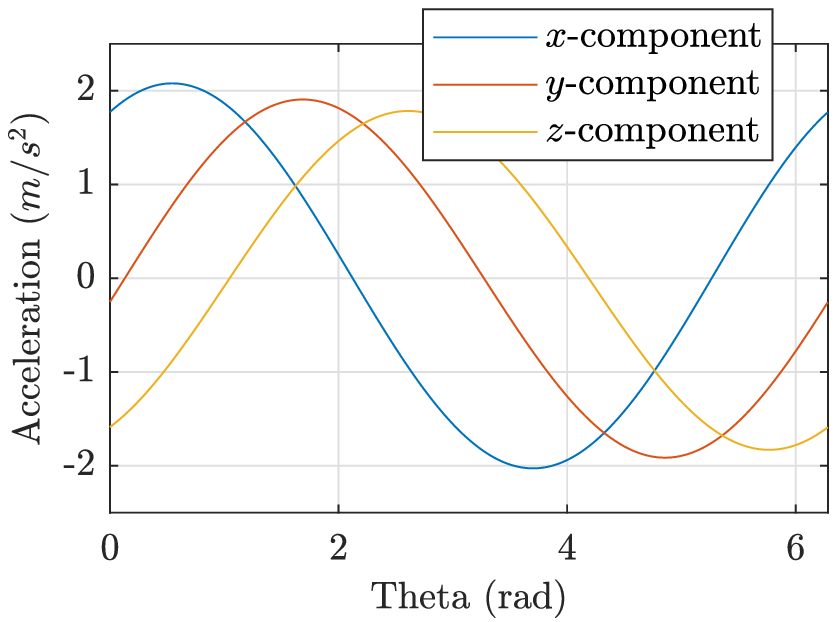}
        \caption{Orbit B acceleration plot.}
        \label{classicalB}
    \end{subfigure}
    \hfill
    \begin{subfigure}[b]{0.49\textwidth}
        \centering
        \includegraphics[width=1\textwidth]{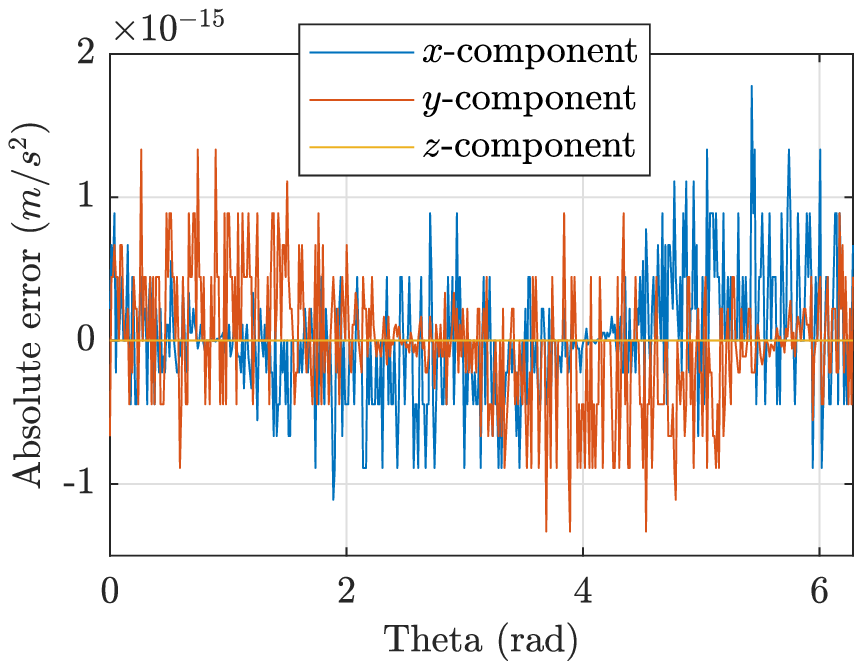}
        \caption{Absolute difference between accelerations produced using PoE and classical methods for orbit A.}
        \label{poeA}
    \end{subfigure}
    \hfill
    \begin{subfigure}[b]{0.49\textwidth}
        \centering
        \includegraphics[width=1\textwidth]{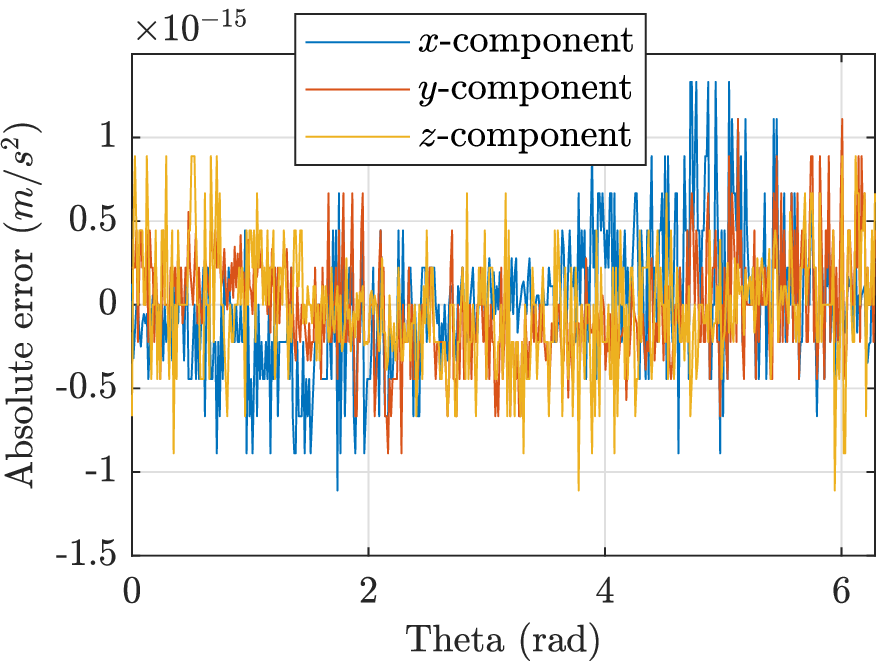}
        \caption{Absolute difference between accelerations produced using PoE and classical methods for orbit B.}
        \label{poeB}
    \end{subfigure}
  \caption{Comparison of classical and PoE accelerations for Orbits A and B.}
\end{figure}

\section{Conclusions and Future Work}

In this work, it was demonstrated that the inertial acceleration of a body can be computed directly from spatial twist, or from spatial acceleration that is obtained by differentiating the spatial twist. The proposed PoE-based method was compared to the classical approach and it was shown that the results are identical. Thus, the demonstrated extension of the PoE-orbital mechanics framework is a successful method for calculating the acceleration of a body moving along an orbit. Now that the inertial acceleration has been derived the PoE method can be used for comprehensive modeling of orbital motion. 

By analytically describing orbital acceleration via PoE, this work opens up a broader class of orbital motion problems to solutions within the PoE framework. 
With this in mind, future work will focus on using the PoE toolbox to describe orbital transfers and relative orbital motion. 
Additionally, practical applications to the orbit determination and maneuver planning problems may be explored. Since all orbital elements can be varied with respect to time, scenarios involving perturbations will be investigated in future works.

\clearpage

\bibliographystyle{AAS_publication}   
\bibliography{main.bib}   

\appendix

\end{document}